# Detecting spatial patterns with the cumulant function.
# Part II: An application to El Niño

Alberto Bernacchia*, Philippe Naveau§, Mathieu Vrac§ and Pascal Yiou§

*: Dipartimento di Fisica, Università "La Sapienza", Roma
§: Laboratoire des Sciences du Climat et de l'Environnement, IPSL-CNRS, France

## Abstract

The spatial coherence of a measured variable (e.g. temperature or pressure) is often studied to determine the regions where this variable varies the most or to find teleconnections, i.e. correlations between specific regions. While usual methods to find spatial patterns, such as Principal Components Analysis (PCA), are constrained by linear symmetries, the dependence of variables such as temperature or pressure at different locations is generally nonlinear. In particular, large deviations from the sample mean are expected to be strongly affected by such nonlinearities. Here we apply a newly developed nonlinear technique (*Maxima of Cumulant Function*, MCF) for the detection of typical spatial patterns that largely deviate from the mean. In order to test the technique and to introduce the methodology, we focus on the El Niño/Southern Oscillation and its spatial patterns. We find nonsymmetric temperature patterns corresponding to El Niño and La Niña, and we compare the results of MCF with other techniques, such as the symmetric solutions of PCA, and the nonsymmetric solutions of Nonlinear PCA (NLPCA). We found that MCF solutions are more reliable than the NLPCA fits, and can capture mixtures of principal components. Finally, we apply Extreme Value Theory on the temporal variations extracted from our methodology. We find that the tails of the distribution of extreme temperatures during La Niña episodes is bounded, while the tail during El Niños is less likely to be bounded. This implies that the mean spatial patterns of the two phases are asymmetric, as well as the behaviour of their extremes.

## 1. Introduction

In geosciences, many datasets consist of multivariate time series (e.g. temperature, precipitation or pressure) measured at different locations. Observations at different places are not independent: they rather display dependencies that cannot be fully understood by simple linear models. Usually, linear correlation is the main investigation tool for such datasets: time series at two locations are taken and the linear correlation is computed. If it is significantly different from zero, it is concluded that there is some dependence in the two time series. This approach is justified because non zero correlations imply dependence. Of course, the converse is not true: insignificant correlations do not imply independence, especially when nonlinearities appear in the dynamics between the variables.

Nonlinearities are usually the rule, rather than the exception, in dynamical processes involved in geosciences. This is especially true in the study of extreme events, i.e. occurrences that largely deviate from the expected behaviour. Using simple linear models, such as linear correlations, does not give accurate results in those cases. Principal Component Analysis (PCA) is the equivalent of correlations in the case of several locations (Rencher 1998): it finds global spatial patterns that are uncorrelated with each other, and calculates for each the corresponding variance. However, if the underlying probability distribution is not Gaussian, uncorrelated patterns are not necessarily independent, and a large variance does not necessarily imply the presence of extreme events.

Here we report an application of a nonlinear method, recently developed by Bernacchia and Naveau (2007), designed to find the spatial patterns responsible for large anomalies. The method is based on the optimization of the cumulant function: beyond the variance, higher order cumulants are taken into account in determining the relevant spatial patterns. In particular, by maximizing the cumulants of the highest accessible order, given the fixed amount of data, the algorithm is able to find the patterns whose projections display the marginal distributions with the fattest tails. As demonstrated in Bernacchia and Naveau (2007), the *Maxima of Cumulant Function* (MCF) are the spatial patterns characterizing large anomalies.

In order to illustrate the methodology of MCF, we focus on the El Niño/Southern Oscillation (ENSO) phenomenon in the Equatorial Pacific. We consider sea-surface temperatures (SST) in the central Pacific, between 1948 and 2005 (see section 2.1: *Equatorial Pacific SST*). ENSO has been the focus of intense research in the last decades, since it dominates the interannual climate signals and has great economical and societal impacts (Philander 1990). It is characterized by large temperature anomalies spanning vast distances across the surface of the entire tropical Pacific Ocean (Wyrtky 1985, Harrison and Larkin 1996). The ENSO phenomenon is related to the highly nonlinear dynamics of the coupled ocean–atmosphere system (Bjerknes 1969, Ghil and Robertson 2000; Neelin et al. 1994), and has a large influence on the global atmospheric circulation (Glantz et al. 1991, Piechota and Dracup 1996, Alexander et al. 2002; Lau and Nath, 2001).

The two anomalous events characterizing ENSO, El Niño and La Niña, are the two extremes of "the Southern Oscillation", but are not exactly symmetric: the warm El Niño phases are generally characterized by a larger magnitude than their cold counterparts La Niña (Burgers and Stephenson 1999; Hoerling et al.1997; Sardeshmukh et al. 2000). The distribution of temperatures is indeed very skewed and far from a Normal distribution. Another indication that the distribution of temperatures is not Normal comes from the observation that El Niño and La Niña anomalies distribute differently: while El Niño is more concentrated in the coast of South America, La Niña is centered in the middle of the Pacific Ocean. However, Nonlinear PCA (NLPCA) was implied in deriving those results (Monahan 2001), whose reliability is under discussion (Christiansen 2005).

In this paper, we find two MCF, i.e. two spatial patterns of temperatures maximizing the cumulant function, we recognize them as El Niño and La Niña, and we confirm their different spatial coherence. We compare the results with other techniques applied to the same ENSO dataset, such as PCA and NLPCA. We show that, under specific assumptions, our algorithm gives more consistent results than PCA, and is more reliable than NLPCA. Finally, we perform univariate Extreme Value analysis on the projections over the two derived spatial patterns. We find a significantly negative shape parameter for La Niña, from which we expect that extremely cold occurrences are characterized by a low temperature bounded tail, while El Niño has a shape parameter close to zero, implying that extremely warm events have unbounded, albeit thin, tails.

## 2. Data and Methods

### 2.1 Equatorial Pacific SST

Data points are monthly anomalies of sea-surface temperatures (SST[1]) over the Equatorial Pacific. Grid points are shown in Fig.1. The area of interest lies between 25S-20N and 150E-280E, with 5 degrees increments. It hence contains 9x27 points, for a total of 243 locations. Data are recorded from January 1948 to December 2005, for a total of 696 months. Data vectors are denoted as $x_t$, where $t$ is time ($t = 1,...,N$), the total number of recordings is $N = 696$. Each vector, a spatial pattern of temperatures, has $n = 243$ components, one for each location. Data vectors $x_t$ are centered on the time average (denoted by angular brackets), i.e.

$$\frac{1}{N}\sum_{t=1}^{N}\overline{x}_t = \langle \overline{x} \rangle = 0$$

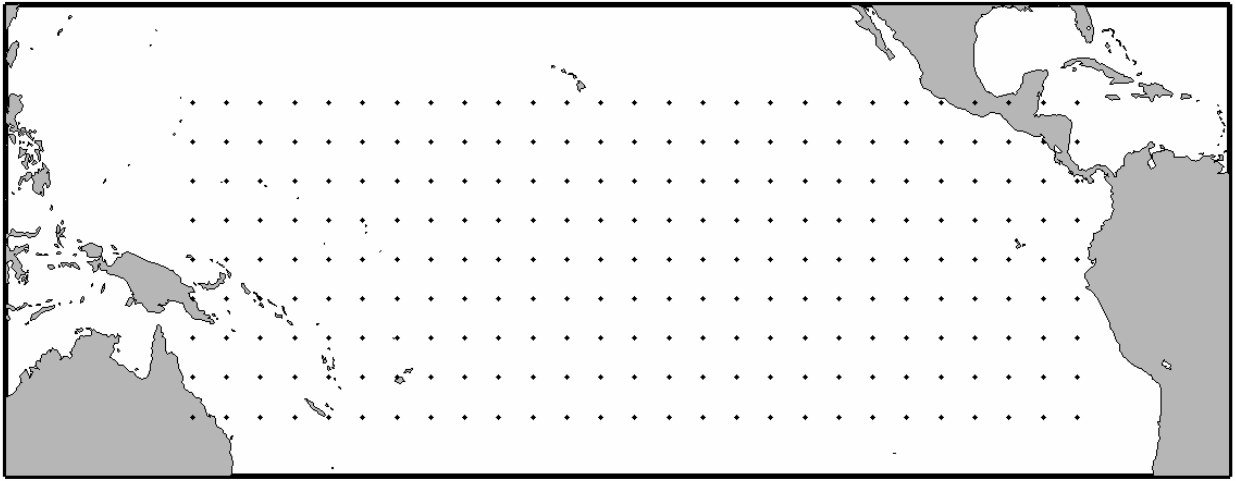

**Figure 1**: Points correspond to locations of data recordings in the central Pacific Ocean ($n = 243$ locations), 5 degrees spaced. Data consist of sea-surface temperatures (SST), monthly anomalies (1948-2005).

### 2.2 The optimizing algorithm

In this section, we briefly summarize the MCF method (for details we refer the reader to Bernacchia and Naveau 2007), and we introduce the methodology for its applications. The method is designed to detect spatial patterns of large anomalies, starting from a dataset composed by time series recorded at various locations, and is based on the optimization of the cumulant function. In Bernacchia and Naveau (2007), the MCF method is described and analyzed in cases in which a probability distribution of data points exists and is known in advance, hence the expression of the cumulant function is known as well. In real applications, of course, the probability distribution (if any exists) is not known, and the cumulant function must be estimated from data.

A spatial pattern, a vector of $n = 243$ components, is denoted as θ, and is normalized to one ($|θ|^2=1$, unit norm). The empirical estimate of the cumulant function is

$$G_s(\bar{\theta}) = \log\left[\frac{1}{N}\sum_{t=1}^{N}e^{s(\bar{x}_t\cdot\bar{\theta})}\right] = \log\left\langle e^{s(\bar{x}\cdot\bar{\theta})}\right\rangle \tag{1}$$

where $x_t$ are the data vectors at different recording times ($t = 1,...,N$), the projection of each data vector $x$ along a pattern $\theta$ is given by the scalar product ($x\cdot\theta$), and $s$ is a positive parameter. In order to introduce the properties of the cumulant function, we show that, for small s, the cumulant function reduce to the variance of data projected along the pattern $\theta$. Using the Taylor expansion of the exponential and the logarithm functions, i.e. $e^s \approx 1+s+s^2/2$, and $\log(1+s) \approx s$, and using the zero mean hypothesis, $<x>=0$, we can rewrite the cumulant function, Eq.(1), as

$$G_s(\bar{\theta}) = \log\left[1 + s\langle\bar{x}\rangle\cdot\bar{\theta} + \frac{s^2}{2}\left\langle(\bar{x}\cdot\bar{\theta})^2\right\rangle\right] = \frac{s^2}{2}\left\langle(\bar{x}\cdot\bar{\theta})^2\right\rangle$$

which is proportional to the variance of data $x$ projected along $\theta$. Our algorithm consists in maximizing the cumulant function with respect to $\theta$, at fixed $s$. Hence, for small s, this corresponds to find the direction $\theta$ along which projected data display maximal variance, whose result is the well known first principal component (PC1) of the data set.

---

[1] Available at: ftp://ftp.cdc.noaa.gov/Datasets/kaplan_sst/sst.mean.anom.nc

In general, for any value of $s$, all powers of s must be taken into account in the Taylor expansion, and the cumulant function is expressed by the following series

$$G_s(\bar{\theta}) = \sum_{i=2}^{\infty} k_i(\bar{\theta})\frac{s^i}{i!}$$

where $k$'s are estimators of the cumulants of data projected along $\theta$. $k_2$ is the variance, $k_3$ is the skewness, $k_4$ is the kurtosis; $k_1$ is the mean and is equal to zero, since data are centered. For a fixed value of $s$, $G$ is a fixed combination of the projected cumulants. If $s$ is small, the combination is dominated by the variance $k_2$, while if the value of $s$ grows, higher and higher order cumulants becomes dominant. If $s$ is large enough, it has been demonstrated (Bernacchia and Naveau 2007) that the patterns $\theta$ maximizing the cumulant function are those whose projection displays the fattest tails in the marginal probability density. These are the patterns of interests.

In order to maximize the cumulant function, an iterative algorithm is used, the iteration is labelled by the index $i$. An initial vector $\theta = \theta_0$ is chosen ($i = 0$), and is updated by the following rule

$$\bar{\theta}_{i+1} - \bar{\theta}_i = \frac{1}{s}\nabla_{\bar{\theta}}G_s(\bar{\theta})$$

where the right hand side (r.h.s.) is proportional to the gradient of the cumulant function with respect to $\theta$, which is rewritten, using Eq.(1), as

$$\bar{\theta}_{i+1} - \bar{\theta}_i = \frac{\frac{1}{N}\sum_{t=1}^{N}\bar{x}_t e^{s(\bar{x}_t\cdot\bar{\theta}_i)}}{\frac{1}{N}\sum_{t=1}^{N}e^{s(\bar{x}_t\cdot\bar{\theta}_i)}} = \frac{\left\langle\bar{x}\, e^{s(\bar{x}\cdot\bar{\theta}_i)}\right\rangle}{\left\langle e^{s(\bar{x}\cdot\bar{\theta}_i)}\right\rangle} \tag{2}$$

This means that the pattern $\theta_i$, during the iteration, moves towards the direction along which the cumulant function has the largest increase. Hence, the algorithm assures that each solution is a local maximum of the cumulant function. The constraint $|\theta|^2=1$ is obeyed by rescaling $\theta$ to unit norm after each iteration step. When the algorithm converges to a stable solution $\theta_{fin}$, the iteration stops and the pattern $\theta_{fin}$ is saved. The algorithm is applied for many different initial conditions $\theta_0$'s, in order to explore all possible solutions in the whole space. Once all the distinct solutions are collected, each normalized (unit norm) $\theta_{fin}$ is multiplied by the standard deviation along the corresponding direction, i.e. $\sqrt{\langle(\bar{x}\cdot\bar{\theta}_{fin})^2\rangle}$, in order to give a scale to the variability of the spatial patterns obtained. The final outcomes, all the $\bar{\theta}_{fin}\sqrt{\langle(\bar{x}\cdot\bar{\theta}_{fin})^2\rangle}$, are the patterns of interests for the given value of *s*.

How to fix the value of the parameter *s*? Our approach is opposite with respect to PCA: we use a value of *s* as large as possible, to maximize cumulants of the highest accessible order, instead of the variance, and to select the patterns for which projected data display the fattest tails. However, for large *s*, due to the finite size of the dataset, the reliability of the estimate of *G* is corrupted by outlier data points, and *s* must be fixed by a tolerance error. For normally and independently distributed data points, with a sample of *N* data vectors, the variance of the estimate of the cumulant function is equal to

$$\varepsilon^2 = \frac{\exp\left[s^2 k_2(\bar{\theta})\right]-1}{N} \qquad (3)$$

If the variance is large, the error in the estimate is large, and the method is expected to be unreliable: different data samples from the same distribution would yield different results. This happens especially for large value of *s*. Our strategy is the following: we fix a tolerance value for the error ε, and we calculate the corresponding value of s, given that we know the size of data sample *N*, and the variance $k_2$. In order to simplify the task, we remove the dependence of the variance on the pattern θ, substituting $k_2$ with its maximum $\sigma_1^2$, that is the variance of the first principal component. Then, we have an upper bound for the error in Eq.(3):

$$\varepsilon^2 \leq \frac{\exp\left[s^2 \sigma_1^2\right]-1}{N}$$

By fixing a tolerance threshold for the error, a critical value is correspondingly fixed for *s*, i.e.

$$s = \frac{\sqrt{\log(1+N\varepsilon^2)}}{\sigma_1} \qquad (4)$$

For the dataset considered here, the maximal variance is $\sigma_1^2 = 42.2$, calculated by standard PCA (the first principal component explains about 50% of the total variance). The error is fixed to the tolerance value ε = 0.1 and, using *N* = 696, we obtain *s* = 0.222. For larger values of s, we expect unreliable estimates of the cumulant function. Note that we used a Gaussian approximation to derive the variance of the estimate: this is not an accurate upper bound for the error if the actual distribution of data points has fatter tails than a Gaussian distribution (see section 4: *Discussion*).

In summary, we fix a tolerance parameter *s*, as large as possible given the size and the variability of the dataset, using Eq.(4), and then we run the algorithm given by Eq.(2), once for each different initial conditions $\theta_0$, finding all the spatial patterns of interests, i.e. those representing the large anomalies. In the present context, other values of *s* where also implemented in the algorithm, in order to investigate the transformation of PC1 to the MCF solutions (see Fig.2), varying *s* from 0 to 0.222. Since the algorithm, Eq.(2), is not well defined for *s*=0, we use the standard PC1 solution in that case.

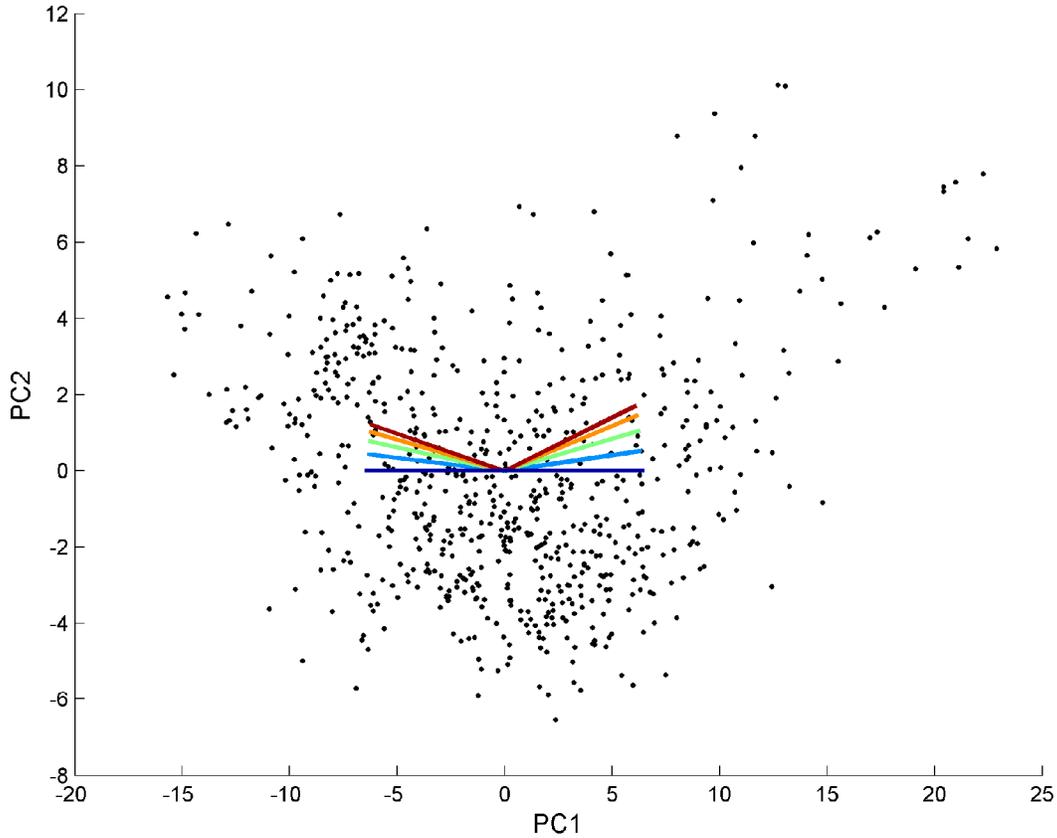

**Figure 2**: Scatter plot of the 696 data points, projected over the first two principal components (PC1 – horizontal, PC2 - vertical). The lines, centered by the sample mean (0,0) correspond to the MCF, for different values of *s*. From bottom to top: *s* = 0 (blue), 0.055 (light blue), 0.111 (green), 0.166 (orange), 0.222 (red). For each value of *s*, two MCF are found, one pointing from the center to the right, the other pointing from the center to the left. For s = 0, the two lines correspond to PC1. For growing values of *s*, the two lines separately move up (V shape), and seem to point towards the two tails of the distribution of data points. The selected solutions, set by the size and the variability of the dataset, are those for *s* = 0.222 (red lines), and are recognized as El Niño (up right) and La Niña patterns (up left).

## *3. Results*

As an implementation of the cumulant function to detect patterns of large anomalies, we consider the ENSO phenomenon in the Pacific Ocean (see section 2.1**:** *Equatorial Pacific SST*). A scatterplot of data vectors is presented in Fig.2. Each vector is represented by one point (*N*=696), where the coordinates on the plot correspond to the scores of the first two principal components (PC1 – horizontal, PC2 - vertical), calculated by standard PCA. We stress that the first two principal components are taken only for illustrative purposes, the following analysis is performed over the full *n*-dimensional space (*n* = 243).

The lines in Fig.2 correspond to the solutions of the algorithm (see section 2.2: *The optimizing algorithm*), i.e. the *Maxima of the Cumulant Function* (MCF), for different values of the parameter *s*. All the lines are centered around the sample mean, (0,0) (data are centered). From bottom to top: *s* = 0 (blue), 0.055 (light blue), 0.111 (green), 0.166 (orange), 0.222 (red). For each value of *s*, we have two MCF solutions, one pointing from the center to the right, the other pointing from the center to the left. For *s* = 0, the two blue lines are horizontal, that is they are parallel to the first principal component. Indeed, the cumulant function reduces to the variance for small $s$, and the first principal component maximizes the variance (since $s = 0$ cannot be set in the algorithm, the solution is derived by standard PCA). For larger values of $s$ the two lines separately move up, one towards the up right part, the other towards the up left part, together displaying a V shape. This result suggests that at least the second principal component, along the vertical, has a significant role when higher order cumulants come into play. According to our interpretation, the two lines for large $s$ (red lines) should point towards the two tails of the distribution of data points, and this seems indeed to be the case for the points of Fig.2, as can be checked by visual inspection.

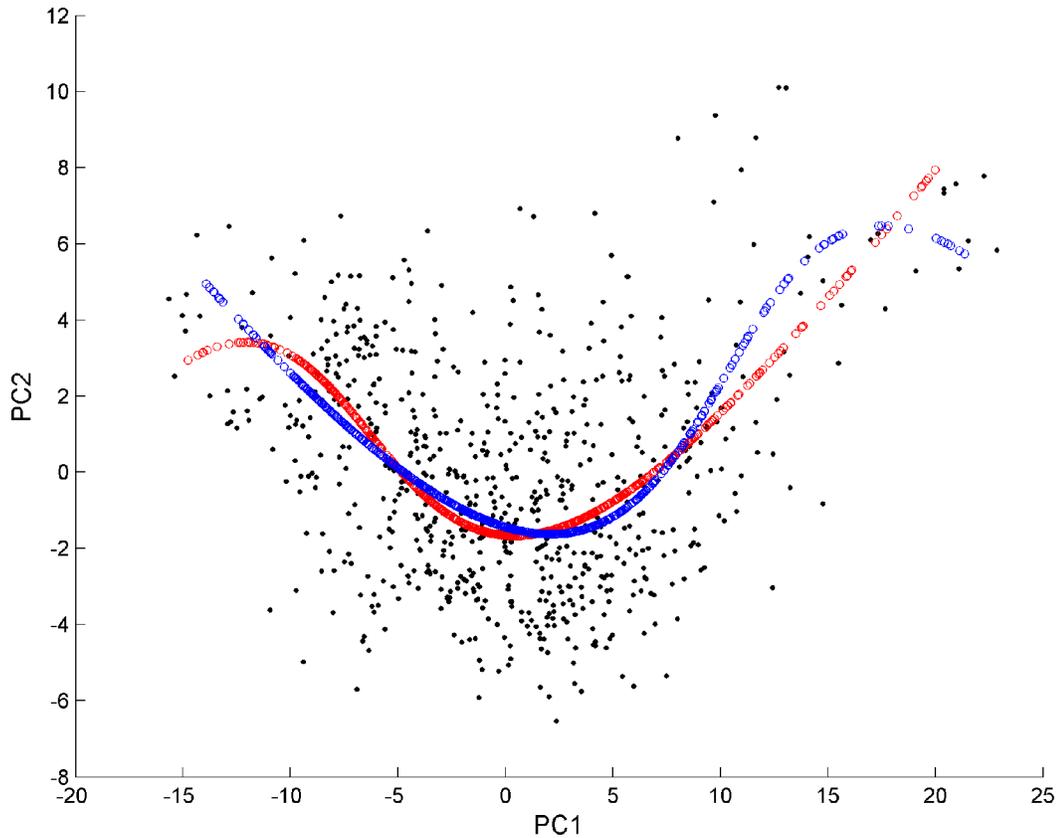

**Figure 3**: Two Nonlinear PCA fits. Data points are the same data of Fig. 2, projected along the first two principal components. The two U-shaped curves (red and blue circles) are two different outcomes of the application of NLPCA: the projections of data points over the two curvilinear NLPCA fits. They resemble the V shape of Fig. 2. Note that the two curves are especially different at the two ends of each curve, and is not clear which of them should be taken as representative for the extremes of data.

For each of the two MCF solutions, differences for subsequent values of $s$ are decreasing with $s$. Hence, for large $s$, the two MCF seem to converge separately to two limit solutions, one up-right and one up-left. Since we want the cumulants of highest possible order to be maximized, corresponding to the largest possible value of $s$, we are tempted to increase $s$ above 0.222 (red lines in Fig.2), in order to reach the limit solutions. However, as explained in section 2.2: *The optimizing algorithm*, the size of the dataset, $N = 696$, limits the extent to which we can calculate reliable estimates with growing $s$, and we fix the maximum at $s = 0.222$. This value of s already selects patterns that are significantly different from the first principal component. Note that the up right tail of data points seems to be longer than the up left tail: this is reflected by the fact that the value of the cumulant function, for each fixed value of s, is always larger for the up right MCF than for the up left MCF (not shown).

For comparison, we present in Fig.3 the fits of Nonlinear PCA (NLPCA). Starting from the same dataset as above, we selected the first three PCs, and we performed ten runs of the NLPCA algorithm (following Hsieh 2004, Monahan 2001). We found two U-shaped solutions (red and blue circles), superimposed in Fig.3 on the scatterplot of the first two PC's, same as Fig.2. They are similar to those described in (Hsieh 2004, Monahan 2001), and they also have some resemblance with the V shape of Fig.2, arising for large values of *s*. The U-shaped curve fits of Fig.3 imply that, at least far from the mass of the distribution, data vectors display prominently positive values of the score of second principal component, supporting our MCF results. However, it is not clear which of the two NLPCA fits, red or blue, we should consider as the good one. Moreover, the difference between the red and blue curves is especially large at the extremes, i.e. at the two ends of the curves, to which we are mainly concerned. Hence, due to the ambiguity of the solutions, NLPCA might not be optimal to investigate the extremes, at least of the present dataset (see section 4: *Discussion* and Christiansen 2005). The MCF method aims at controlling the reliability of the results by improving the consistency of the pattern estimates.

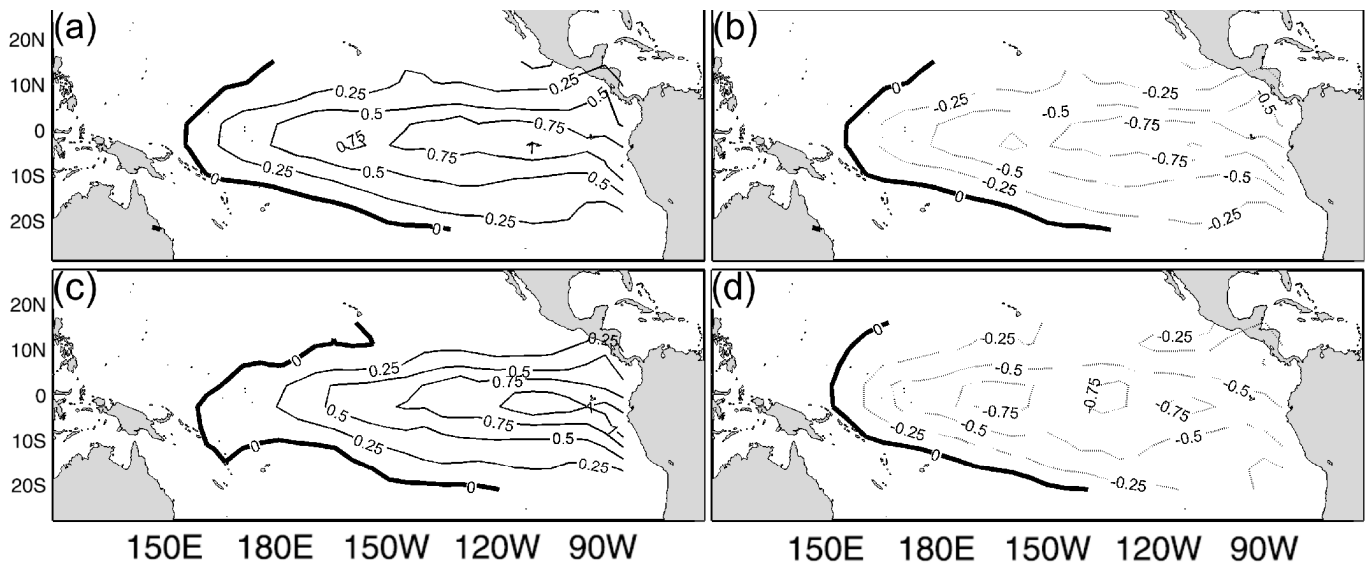

**Figure 4**: (a),(b): The first principal component and its opposite. (c),(d): The two MCF, recognized as El Niño (c) and La Niña (d). They correspond to the lines in Fig.2, up right – El Niño, up left – La Niña. Note that (a) is similar to (c), but (c) is characterized by stronger positive anomalies at the coast of South America, and weaker positive anomalies in the central Pacific. Similarly, (b) resembles (d), but (d) has stronger negative anomalies in the central Pacific and weaker negative anomalies at the coast of South America.

As the final outcome of our procedure, we select the MCF for *s* = 0.222 (red lines in Fig.2), each of the two corresponds to a spatial pattern of temperatures. They are the spatial patterns representative for large anomalies, and are presented in Fig.4c,d. We recognize them as El Niño (c) and La Niña (d) patterns: the former is characterized by strong positive anomalies near the coast of South America, while the latter has negative anomalies over the central Pacific. In Fig.2, the El Niño pattern corresponds to the long up-right tail of data points, while the La Niña pattern corresponds to the shorter up-left tail. For comparison, we present in Fig.5 the SST anomalies patterns during three El Niño (January 1983, 1992, 1998) and three La Niña events (December 1950, January 1974, September 1988).

The spatial pattern corresponding to the first principal component (PC1), and its opposite, are presented respectively in Fig.4a,b. Note that the first PC (4a) resembles El Niño (4c), and the negative first PC (4b) resembles La Niña (4d). However, their difference is significant, and is plotted respectively in Fig.6b,c. El Niño is characterized by stronger positive anomalies at the coast of South America, and weaker positive anomalies in the central Pacific, while La Niña is characterized by stronger negative anomalies in the central Pacific and weaker negative anomalies at the coast of South America. This asymmetry between El Nino and La Nina is also noticeable in the six examples of Fig.5, even if in some cases (January 1992 – September 1988) the asymmetry is less evident. The second principal component (PC2) is plotted in Fig.6a. Note that the difference between El Niño with respect to PC1, in Fig 6b, and the difference between La Niña and the negative PC1, in Fig.6c, both have a structure very similar to the PC2, Fig.6a. This confirms the results illustrated in Fig.2, for which PC2 seems to play a significant role for large anomalies. In particular, PC2 gives a positive contribution to both El Niño and La Niña: they have respectively 27% and 19% overlaps with PC2 (in terms of scalar product).

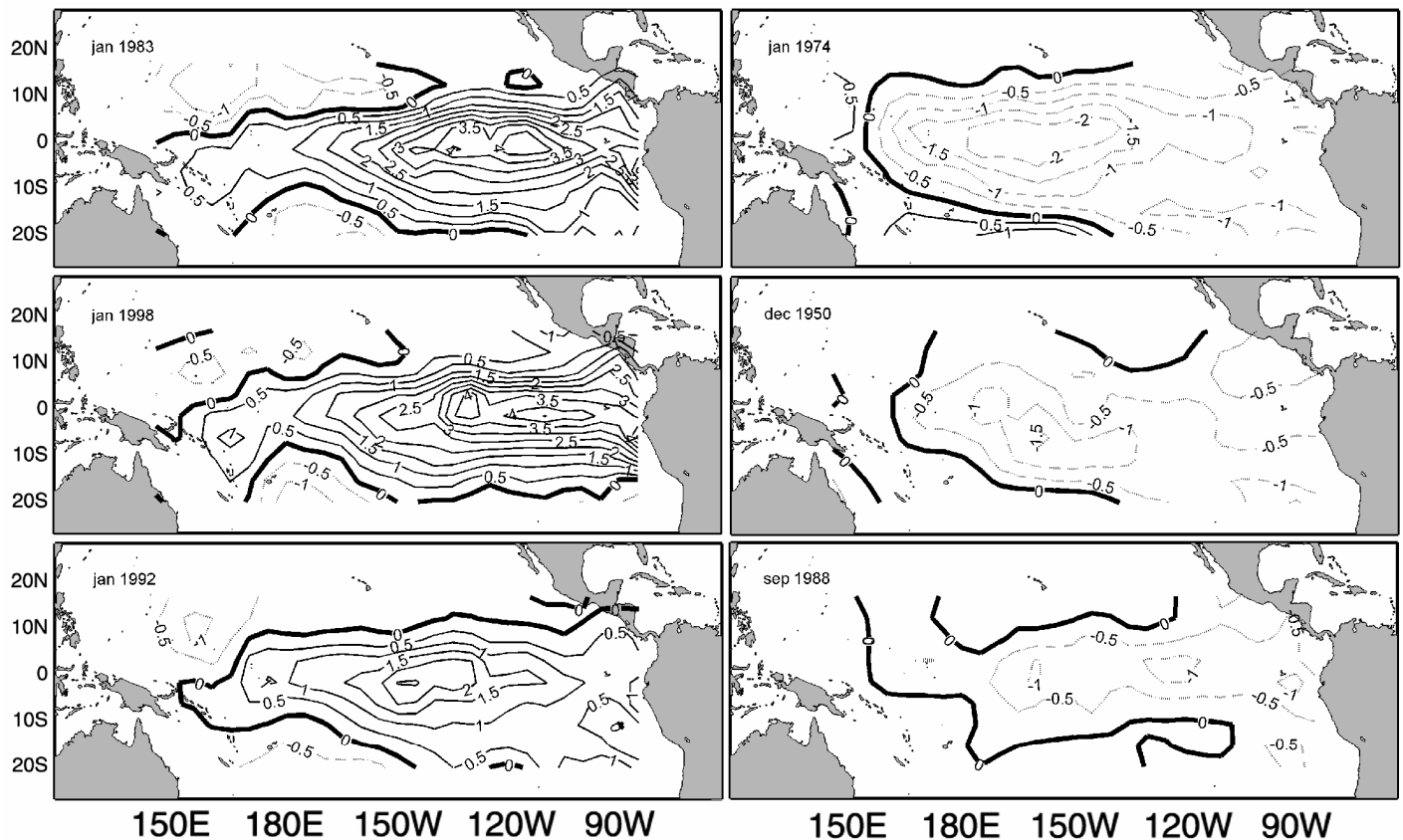

**Figure 5**: Six examples of El Nino and La Nina events. Three El Nino events (January 1983, 1992, 1998) and three La Niña events (December 1950, January 1974, September 1988). Note that the asymmetry between the two patterns is partially captured by the MCF solutions, Fig.4c,d.
9

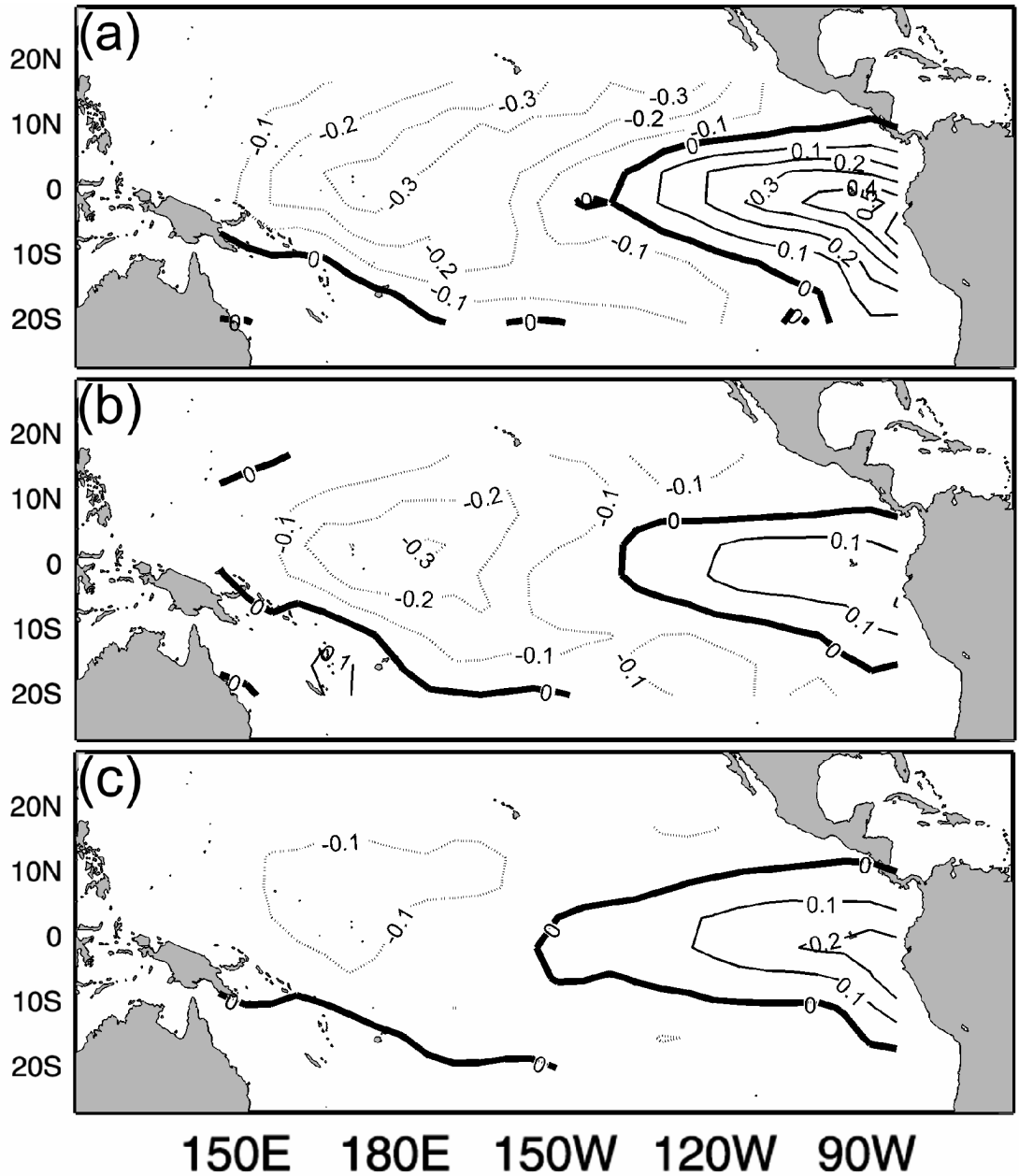

**Figure 6**: (a) The second principal component. (b) The difference between the MCF El Niño pattern (Fig.4c) and the first principal component (Fig.4a). (c) The difference between the MCF La Niña pattern (Fig.4d) and the negative first principal component (Fig.4b). Note the similarity in the spatial structure of both pattern differences (b),(c), respect to the second principal component (a), indicating that the latter is the main source of the asymmetry between the El Niño and La Niña patterns.

From the above analysis, we conclude that El Niño and La Niña patterns are non-symmetric, i.e. one is not just the opposite sign of the other: positive and negative anomalies are displaced over separated regions, and the displacement is mostly controlled by the second principal component. The same kind of asymmetry was found using NLPCA (Hsieh 2004, Monahan 2001), repeated here for the dataset considered and illustrated in Fig.3. Again, the second principal component was found to play a significant role in determining the asymmetry. However, we showed that a naïve application of NLPCA can give ambiguous results, especially concerning the extremes. Moreover, due to its computational expense, the NLPCA fit was performed in the space of the first three PC's, while MCF are derived on the whole 243-dimensional space, hence the latter does not rule out contributions from other principal components. Both methods, MCF and NLPCA, are able to find nonsymmetric patterns, since they both deal with nonlinearities, but the cumulant function is more reliable for the purpose of detection of large anomalies. While NLPCA tries to keep the whole structure of data at all scales, MCF concentrates on tails, and gains in simplicity: it is parameter-free and unambiguous.

Since the MCF solutions are found in the form of vectors in the space of data, the projections of data points along these vectors are easily computed, and the time series of the spatial patterns corresponding to El Niño and La Niña are separately studied with standard Extreme Value Analysis (Coles 2001). Annual maxima are extracted from the two resulting time series, and are fitted with a Generalized Extreme Value (GEV) distribution (note that the cold La Nina is analysed by maxima, not of temperatures, but of its projection). GEV distributions depend on three parameters: a location parameter $\mu$, a scale parameter $\sigma$, and a shape parameter $\xi$ (Coles, 2001). The location and scale parameters, respectively, indicate approximately the peak and the width of the GEV distribution, while the shape parameter gives an indication on the tail of the distribution (i.e. short, light or heavy tail). All the parameters are estimated by maximizing a likelihood function from which we retrieve confidence intervals. We found that while the location and the scale parameters are quite similar in the El Niño and La Niña cases ($\mu_{Niño}$ is 25% smaller than $\mu_{Niña}$, and $\sigma_{Niño}$ is 10% larger than $\sigma_{Niña}$), the shape parameter shows significant differences.

GEV fits of annual maxima are plotted in Fig.7 (top), of respectively El Niño (left) and La Niña (right) projections, using rescaled variables ($Z = (X-\mu)/\sigma$). The corresponding Quantile-quantile (QQ) plots are given in Fig.7 (center), to indicate the goodness of fit of the GEV representation of the extremes. Annual maxima of La Niña projections are fitted by a significantly negative shape parameter ($\xi = -0.195 \pm 0.108$), indicating that the marginal density tail of data projected along La Niña is bounded. Conversely, for El Niño projections, the shape parameter of annual maxima is undistinguishable from 0 ($-0.039 \pm 0.102$) indicating a light and potentially unbounded tail. Return periods of extremes are plotted in Fig.7 (bottom) respectively for El Niño (left) and La Niña (right), within 1SD confidence interval (dashed lines). The six examples of Fig.5 are denoted by red and blue dots, respectively for the three El Niño and the three La Niña events. For a fixed value of $z$, return periods of El Niño events are smaller then La Niña events. Moreover, El Niño events for large values of $z$ are more likely to appear than their La Niña counterparts. These results indicate that the tail of El Niño projections is longer than the tail of La Niña projections. In summary, according to our interpretations of El Niño and La Niña spatial patterns, while we expect that La Niña cold extreme events are confined by a lower temperature bound, similar conclusions cannot be drawn for El Niño, for which an upper bound in extremely warm events is not guaranteed.

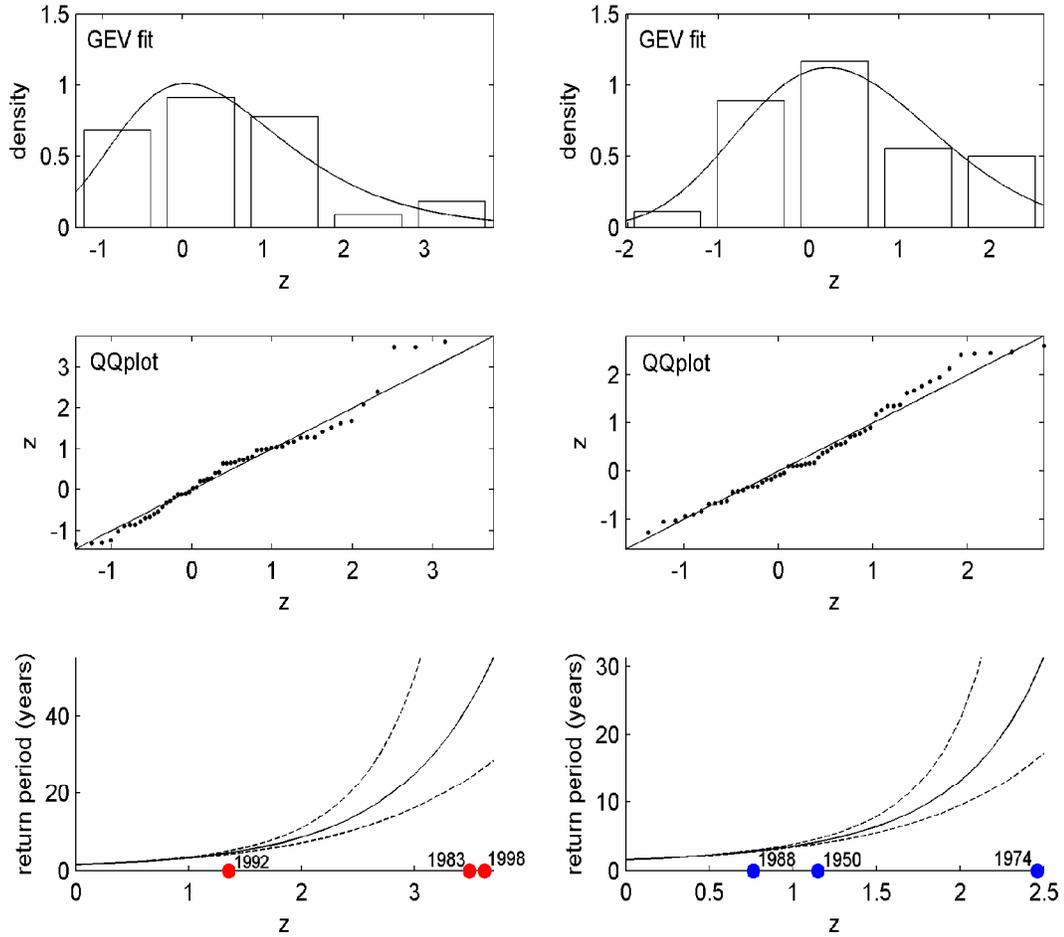

**Figure 7**: : (Top): Generalized Extreme Value fits of the annual maxima of respectively El Niño (left) and La Niña (right) projections. (Center): The corresponding QQplots, indicating the goodness of the fit. (Bottom): Return periods of annual maxima, within a 1SD confidence interval (dashed lines), for El Niño (left) and La Niña (right). Red dots and blue dots correspond respectively to the three El Niño and the three La Niña events presented in Fig.5.

## *4. Discussion*

In this paper, we have introduced a novel methodology for selecting the spatial patterns representative for large deviations in the dataset (Bernacchia and Naveau 2007). It consists of finding the vectors for which the multivariate cumulant function, estimated from data, is maximal (*Maxima of Cumulant Function*, MCF). We chose to apply this algorithm to temperature variability in the equatorial Pacific, which is controlled by ENSO. We found two MCF solutions and we identified them as the spatial patterns of El Niño and La Niña. As in the case of Principal Component Analysis (PCA), these spatial patterns are found in the form of directions in the space of data. However, while PCA concerns the mass of the distribution, the MCF cares about large deviations. In both cases, the subspaces spanned by different patterns are ordered: in PCA the order follows the fraction of variance present in each subspace, while in our algorithm the order is given by the value of the cumulant function for each MCF solution.

The MCF spatial patterns of El Niño and La Niña are asymmetric: while the pattern of the warm El Niño is concentrated over the west coast of South America, the cold La Niña is centered in the middle of the Pacific Ocean. Indeed, the solutions of the MCF algorithm are not expected to be symmetric: this is considered as an advantage with respect to PCA, since large anomalous patterns are not expected to be, in general, neither parallel nor orthogonal. This is the case for the ENSO dataset: the first principal component has a significant overlap with both El Niño and La Niña MCF, with different signs. Then, one is tempted to recognize the positive part of the first principal component as El Niño, and the negative part as La Niña. However, we have shown here that the MCF depart from the first principal component, and at least the second principal component play a role in determining large anomalies. Other methods that allow oblique solutions, such as VariMax and related techniques (Rencher 1998), are not supposed to detect patterns of large anomalies.

A similar asymmetry between El Niño and La Niña was reported using NLPCA (Hsieh 2001, Monahan 2001), but this does not support the use of NLPCA for science. We have shown here that application of NLPCA does not give consistent results, especially concerning the extremes. A criticism of NLPCA, regarding its applications to atmospheric circulation, was also reported in (Christiansen 2005). In our opinion, NLPCA is not adapted for this type of dataset and cannot be used to interpret unusually large deviations, since results depend on the ambiguous choice of the parameters and on the accessibility of the global solution. Here, we have indeed found two different solutions with the same set of parameters.

Once the MCF spatial patterns of El Niño and La Niña have been found, a linear projection is performed, with an easy geometrical interpretation, and Extreme Value Analysis applied, to determine the anomalous behaviour of each projected time series separately. While La Niña projection is found to have a negative shape parameter, corresponding to a finite bound for the extremes, the El Niño projection has around zero shape parameter, indicating a possible unbounded tail. According to the larger value of the cumulant function for El Niño with respect to La Niña, the former is characterized indeed by a fatter tail.

Note that maximizing the cumulant function is computationally very cheap, has no free parameter, and has the advantage of searching for local solutions (two in the present case), all of which are of interest. Even if it is illustrative to check out the solutions for different values of $s$, it is not a free parameter: $s$ is fixed by a tolerance error $\varepsilon$ in the estimate of the cumulant function from raw data (here, we set $\varepsilon = 0.1$, and we get $s = 0.222$). When a local maximum of the cumulant function is found, it is always a good solution. Instead, when a solution of NLPCA is found, it must be questioned if it is the global or just a local solution.

Since the tolerance value of $s$ depends on the size $N$ of the sample, the MCF are biased estimators of the asymptotic solutions, i.e. the hypothetical vectors obtained with an infinite sample and an infinite $s$. However, the limit is assumed to be consistent and to be reached quite rapidly. The method is expected to be appropriate in cases in which the density of data points is not markedly multimodal and decays not less than exponentially fast with the distance from the center of mass of the distribution. In general, for a given size of the sample, fatter tails correspond to larger errors, because single outlier data points becomes dominant the estimate of $G$. If the density decays slowly, for instance as a power law, the method is expected to give inaccurate results.

In (Bernacchia and Naveau 2007), the MCF method has been found to give appropriate results, and has been solved analytically for three model cases. In the special case of normally distributed data, the MCF corresponds to the first principal component for all values of $s$. In that case, all the information about large deviations is contained in the covariance, whose structure is revealed by the principal components. Another special case arises when the covariance matrix has all equal variances (eigenvalues): in that case MCF reduces to Independent Component Analysis (ICA, see Hyvarinen 2000), i.e. it finds the independent components (if any exists) of the dataset. In general, for any shape of the distribution of data points, MCF is able to find the projections displaying the locally largest tails.

## *Acknowledgements*

This work was supported by the european E2-C2 grant. We would like to thank Isabella Bordi, Marcello Petitta and Alfonso Sutera for valuable discussions.